\documentclass[preprint, authoryear, 12pt]{elsarticle}
\usepackage{amssymb}
\usepackage{amsthm}
\usepackage{setspace}
\usepackage{graphicx}
\usepackage{lineno}
\journal{Planetary and Space Sciences}
\begin{document}
\begin{frontmatter}
\title{Models of dust around Europa and Ganymede}
\label{1}
\label{2}
\author[1]{K. Miljkovi\'c\fnref{cor1}}
\author[2]{J. K. Hillier}
\author[1]{N. J. Mason}
\author[1]{J. C. Zarnecki}
\address[1]{Department of Physical Sciences, The Open University, Walton Hall, Milton Keynes, MK7 6AA, United Kingdom}
\address[2]{Institut f\"ur Geowissenschaften, Universit\"at Heidelberg, 69120 Heidelberg, Germany}
\fntext[cor1]{Corresponding author address: Institute de Physique du Globe de Paris, 4 avenue de Neptune, 94100 Saint-Maur-les-Foss\'es, France}
\newpage
\begin{abstract}
We use numerical models, supported by our laboratory data, to predict the dust densities of ejecta outflux at any altitude within the Hill spheres of Europa and Ganymede. The ejecta are created by micrometeoroid bombardment and five different dust populations are investigated as sources of dust around the moons. The impacting dust flux (influx) causes the ejection of a certain amount of surface material (outflux). The outflux populates the space around the moons, where a part of the ejecta escapes and the rest falls back to the surface. These models were validated against existing Galileo DDS (Dust Detector System) data collected during Europa and Ganymede flybys. Uncertainties of the input parameters and their effects on the model outcome are also included. The results of this model are important for future missions to Europa and Ganymede, such as JUICE (JUpiter ICy moon Explorer), recently selected as ESA's next large space mission to be launched in 2022.
\end{abstract}
\end{frontmatter}
\newpage
\section{Introduction.}
Micrometeoroids (solid micron-sized dust particles) are a common constituent of the Solar System. They can easily reach the surfaces of atmosphereless bodies and, upon impact, cause the ejection of surface material into the surrounding space. The ejected dust fragments populate the space around the host bodies, where a part of the ejecta escapes and the rest eventually falls back to the surface. This process can be numerically modelled. In this paper we investigate dust around Europa and Ganymede created by micrometeoroid bombardment. \\*
Our study is built upon the models developed by \citet{Krivov2003} and \citet{Kruger2003} supported by our impact experiments \citep{Miljkovic2011a} and hydrocode impact modelling to reduce the number of variables in the model. The impact experiments were made using the light gas gun (LGG) at the Open University's Hypervelocity Impact (HVI) laboratory \citep{Miljkovic2011a, Patel2010, McDonnell2006, Taylor2006}. A series of high velocity impacts were made at $2\,km\,s^{-1}$ using 1 mm diameter stainless steel balls as projectiles into pure water ice and sulphate hydrated mineral targets. The ejecta size \citep{Miljkovic2011a} and velocity distributions were measured and subsequently modelled using ANSYS AUTODYN finite element hydro-dynamic shock physics code \citep{Miljkovic2010, Pierazzo2010}. \\*
The dust cloud model presented in this paper characterizes the dust environment (size, density, flux and velocity distribution of such dust) around Europa or Ganymede, predicts the dust densities of the ejecta outflux at any altitude within Europa's and Ganymede's Hill spheres (radius of $13300\,km$ or $8.5\,R_{E}$, where $R_{E}=1565\,km$ for Europa and $32000\,km$ or $12\,R_{G}$, $R_{G}=2631\,km$ for Ganymede) and can evaluate the dust density at any altitude as a function of the size distribution of the dust. We choose a discrete selection of altitudes and dust masses to give a representative set of results. \\*
Our results are important for future space missions to Jupiter System that carry a dust detector onboard. This study can be further applied to estimate the dust counts into a   dust detector in orbit around Europa and/or Ganymede. A dust detector has been proposed as part of a payload for a space mission to Europa and Ganymede. Initially named Laplace in 2007, it was renamed the Europa-Jupiter System Mission after ESA and NASA joined proposals in 2009 \citep{Blanc2009} for a major mission to Jupiter System. In 2011, EJSM was reformed again into an ESA-led mission to Ganymede with flybys to Europa and Calisto, named JUICE (JUpiter ICy moon Explorer) \citep{Dougherty2011}. JUICE has recently been selected as ESA's next large mission for launch in 2022. If a dust detector is included in the payload, an in-situ analysis of the dust that surrounds Europa and Ganymede will be possible, which could provide information about the surface, as its composition should be "written" in the detected dust \citep{Miljkovic2011b}. \\* 
The proposed dust detector should not only be capable of determining the density of dust in the cloud, but may provide chemical analysis of captured dust \citep{Miljkovic2008}, as was the case with the analysis of the Jovian Dust Stream particles by Cassini's CDA (Cosmic Dust Analyser) \citep{Postberg2006}. Chemical abundance maps of Europa and Ganymede even at low spatial mapping resolution show that water ice is non-uniformly distributed over the surfaces of Europa and Ganymede. The Galileo NIMS (Near Infra-red Mapping Spectrometer) spectra identified these impurities as hydrated minerals, sulphates and possibly hydrocarbons \citep{McCord1998}. As the surface material is ejected by micrometeoroid bombardment, it can be expected that the dust particles around Europa will be composed of water ice, sulphur salts and their decomposition products, including any potential organic compounds \citep{Miljkovic2011b}. It should be emphasized that a dust detector has never visited Europa or Ganymede at such a close orbital distance or spent a longer time than in flybys. No chemical analysis of the dust in Jupiter System has been made yet, apart from Cassini's Cosmic Dust Analyser (CDA) measurements of the Io stream dust at more than $1\,AU$ away from Jupiter \citep{Postberg2006}. 

\subsection{Comparison of the present dust cloud model with previous models.}
In previous work, \citet{Krivov2003} developed a spherically symmetric case for an atmosphereless body with applications to Ganymede and the Saturnian satellites to help the interpretation of Cassini's CDA (Cosmic Dust Analyser) measurements; \citet{Kruger2000} developed a dust cloud model around Ganymede to explain the Galileo DDS measurements; \citet{Sremcevic2003} investigated the asymmetry of the dust clouds around Galilean and Saturnian satellites; \citet{Sremcevic2005} compared the Galileo data with their models for Europa, Ganymede and Callisto, whereas \citet{Kruger2003} investigated the dust clouds around all four Galilean satellites. \\*
The differences between the dust model presented here and that of \citet{Kruger2003} are the following: their slope of the cumulative ejecta mass distribution was approximated, whereas in our model our slope was derived directly from the impact experiments published in \citet{Miljkovic2011a}. \citet{Kruger2003} reported that dust have mean mass of $10^{-11}$g around Europa and $10^{-13}$g around Ganymede, whereas in our model the ejecta fragment size distributions were calculated, and were in range between $10^{-15}$g-$10^{-7}$g. Ejecta speed distributions in \citet{Kruger2003} were represented as the distribution of ejecta material having speeds higher than a certain speed, that is dependent on the minimum ejecta speed of fragments and the power-law slope of the distribution, which were unknown variables fitted to match the Galileo data, whereas in our model, a size-velocity relation was applied to all the ejected fragments in order to have more precise outflux and spatial densities of ejected dust. We primarily focus on the short-lived, bound ejecta at a distance from the surface at which a spacecraft may orbit. Therefore, any asymmetry effects in the spatial density of ejected fragments caused by Europa's orbital motion can be excluded (such were considered by \citet{Sremcevic2003}) as well as the charging of dust fragments.

\section{Micrometeoroid influx into Europa's and Ganymede's surface.}
In the Jovian system, there are five distinct sources of dust around Europa and Ganymede. These are: (i) asteroidal and (ii) halo dust populations, as part of the interplanetary dust particle (IDP) population distinguished by their location and not necessarily by their origin \citep{Divine1993}; (iii) interstellar dust (ISD) that originate from beyond the Solar System; (iv) the Io stream and (v) ring dust that from the Jovian system itself. 

\subsection{Influx into Europa's and Ganymede's surface from the Solar System and interstellar region.}
\begin{figure}[htb]
\begin{center}
\includegraphics[scale=0.85]{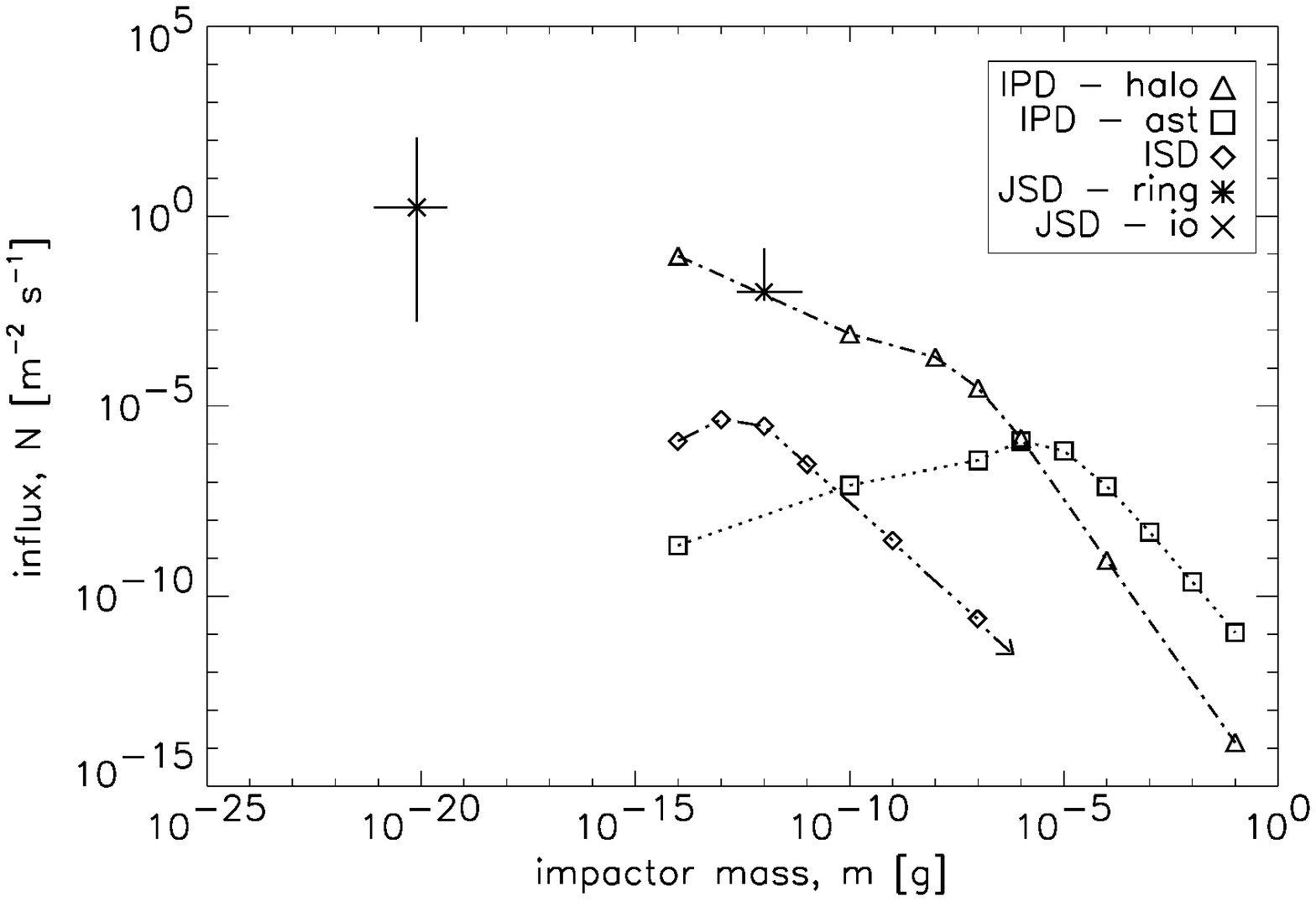}
\end{center}
\caption{Total impact flux (influx) onto the surfaces of Galilean satellites showing all dust populations considered as a source of dust around Europa and Ganymede. The interplanetary asteroidal (IDP-ast), halo (IDP-halo) and interstellar dust (ISD) populations are presented as mass distributions \citep{Divine1993, Landgraf2000}. Due to a lack of data on the Jupiter system dust (JSD), Io (JSD-Io) and ring (JSD-ring) dust are presented as single points with estimated uncertainties \citep{Kruger2004, Krivov2002}. The total influx includes the gravitational and magnetic field focusing and deflection of fragments smaller than $10^{-15}$g by the magnetic field from outside the Jovian system \citep{ColwellandHoranyi1996}.}
\label{dust} 
\end{figure}

\citet{Divine1993} created a phase density model to predict micrometeoroid fluxes at different distances from the Sun. According to this model, there are five distinct interplanetary dust populations, out of which only two (asteroidal and halo) are present at Jupiter' s orbital (heliocentric) distance, $r$. The cumulative number of IDPs per unit volume (spatial density, $N_{M}$) whose mass exceeds $m$ can be calculated as a function of inclination, $i$, represented by elliptical latitude $\lambda$, eccentricity, $e$, and the IDP mass represented by $H_{m}$ is shown in Eq. \ref{divine} \citep{Divine1993}. 
\begin{equation}
\label{divine}
N_{M}=\frac{1}{\pi}\int_0^\infty H_{m}\,dm \int_0^{\pi/2} N_{1}\,d\chi \int_{e_{\chi}}^1 \frac{p_{e}\,de}{\sqrt{e-e_{\chi}}} \int_{\mid\lambda\mid}^{\pi-{\mid\lambda\mid}} \frac{p_{i}\sin{i}\,di}{\sqrt{(\cos{\lambda})^2 - (\cos{i})^2}}
\end{equation}
The mass distribution of micrometeoroids $H_{m}$ is independent of the position and velocity of dust particles in the Solar System, the cumulative mass distribution of dust particles $N_{1}$ is a function of radial distance from the Sun; $\chi=\sin^{-1}({r_{1}/r})$ and $e_{\chi}=(r-r_{1})/(r+r_{1})$, where $r_{1}$ is the perihelion distance, $r$=5.2 AU and $\lambda=0^{o}$ are Jovian heliocentric distance and the equatorial latitude, $p_{e}$ and $p_{i}$ are normalized eccentricity and latitude distributions. Discrete values for $N_{1}$, $H_{m}$, $p_{e}$ and $p_{i}$ were taken from \citet{Divine1993} for the respective asteroidal and halo populations and integrated in Eq. \ref{divine}. $H_{m}$ was integrated over mass bins ($\Delta{m}$), in order to transform the continuous mass distribution into a binned one (Eq. \ref{binnedN}).
\begin{equation}
\label{binnedN}
N_{M}=const \int_m^{m+\Delta{m}} H_{m}\,dm
\end{equation}
In newer meteoroid codes, the product of functions $N_{1}$x$p_{e}$x$p_{i}$ is replaced by a single vector function, which provides a more accurate and detailed dust flux calculation (e.g. \citet{Dikarev2005}). However, due to lack of observational data at 5 AU from the Sun, and since later meteoroid models were based on \citet{Divine1993} (e.g. \citet{Jehn2000, Staubach1997}), we consider the model by \citet{Divine1993} to be satisfactory for a preliminary dust flux calculation at Jupiter's distance for dust coming from asteroidal and halo sources. It should be noted that we have neglected Solar radiation pressure due to Jupiter's large distance from the Sun. \\*
\citet{ColwellandHoranyi1996} calculated that at 100 $R_{J}$ away from Jupiter, the Oort Cloud dust (highly inclined and eccentric dust associated with Divine's halo population) shown as triangles in Fig. \ref{dust} and planetary dust (low inclination and low eccentricity orbits, associated with Divine's asteroidal population) shown in squares in Fig. \ref{dust}, move at a mean speed of 23.6 $kms^{-1}$ and 6.6 $kms^{-1}$, respectively. \\*
The interstellar dust (ISD) can be observed far above the equatorial plane. The Ulysses spacecraft monitored the ISD activity at high ecliptic latitudes between 3 and 5 AU from the Sun, which was far away from contamination by IPD \citep{Grun1997}. ISD penetrates the solar system at about 26 $kms^{-1}$ \citep{Landgraf2000, Kruger2007}. Fig. \ref{dust} shows the ISD flux data taken by Ulysses spacecraft (that measured grain masses between $10^{-11}$ g and $10^{-7}$ g) and ground based radar meteor observations made by AMOR (Advanced Meteor Orbit Radar) facility, that measured the flux of ISD dust larger than $10^{-7}$ g. AMOR data in Fig. \ref{dust}. is shown as a tail with a slope of -1.1 \citep{Landgraf2000}. Fig. \ref{dust} also shows that the ISD flux is lower than the IDP flux. Both the IDP and ISD densities were approximated to 2.5 $gcm^{-3}$.\\*
All grains approaching Jupiter are exposed to the Jovian magnetospheric plasma and magnetic field with which they interact. The effect of the Lorentz force on grain motion depends on the grain charge to mass ratio \citep{Horanyi1993a}. The smaller the grain, the stronger is the perturbation, for a given charge. In order to account for the effects of the Jovian magnetosphere, the influx dust masses in the range $10^{-14}$-$10^{-10}$ g were multiplied by a factor of 2.65, which is an averaged magnification value from the calculations made by \citet{ColwellandHoranyi1996}. Jovian gravitational focusing increases the speed of approaching micrometeoroids and therefore the spatial density of micrometeoroids becomes higher closer to the planet. Grains smaller than $10^{-15}$ g were excluded from our model as they are likely to become deflected from entering Jupiter's magnetic field \citep{ColwellandHoranyi1996}. On grains larger than $10^{-10}$g gravitational effects are more dominant. \\*
The gravitationally focused influxes at Europa's and Ganymede's distances from Jupiter, $F_{in}$, were calculated using Eq. \ref{Krivov} \citep{Spahn2006}.
\begin{equation}
\label{Krivov}
F_{in}/F_{in}^{\infty}=0.5 \sqrt{1+T}[\sqrt{1+T}+1+ \sqrt{1+T-(R_{p}/a)^2(1+Ta/R_{p})}]
\end{equation}
$F_{in}^{\infty}=N_M \upsilon_{imp}^{\infty}$ is the non-focused micrometeoroid flux at Jupiter's heliocentric distance, for all different dust populations. $T=2GM_{p}/(a(\upsilon_{imp}^{\infty})^2)$, $\upsilon_{imp}^{\infty}$ is the micrometeoroid non-focused velocity, $\upsilon_{imp}$ is the micrometeoroid gravitationally focused velocity, $G$ is the gravitational constant, $M_p$ is Jupiter's mass, a is the mean semi-major axis of Europa's or Ganymede's orbit and $R_p$ is Jupiter's radius.
\subsection{The influx onto Europa and Ganymede sourced within the Jovian system.}
There are also dust populations that originate from inside the Jovian system and intersect the orbits of Europa and Ganymede (Fig. \ref{dust}). Io stream dust (5-15 $nm$ in size, assuming the dust is spherical \citep{Kruger2004}, 1.5-2.16 $gcm^{-3}$ in density \citep{Kruger2004, Postberg2006} and spatially averaged dust density between $10^{-3}$ and $10^{-8}$ $m^{-3}$ (\citet{Kruger2004}) and ring dust, 500-1000 $km^{-3}$ in spatial dust density \citep{Kruger2004, Krivov2002}, 0.6-2 $\mu{m}$ in size, assuming spherical shape and approximately 2 $gcm^{-3}$ in bulk density \citep{Kruger2004} also contribute to the bombardment of Europa and Ganymede. Io dust quickly becomes charged and its motion is then influenced by the Jovian and interplanetary magnetic field \citep{Flandes2011}, forming a so-called ballerina skirt shaped dust stream and moving at the co-rotational plasma velocity \citep{Horanyi1993b}. Therefore it can be assumed that the Io stream dust approaches Europa at a speed of about 120 $kms^{-1}$ and Ganymede at about 190 $kms^{-1}$. The dust rings are composed of dust grains moving in prograde and retrograde direction. Prograde dust is likely to have been ejected from the surfaces of the Galilean moons and assumed to be four times more abundant \citep{Thiessenhusen2000} than retrograde dust. Retrograde dust is likely to be populated by captured IDPs. It is not known if non-gravitational processes affect the motion of the dust in the rings \citep{Krivov2002}. Europa and Ganymede move together with the prograde dust ring at approximately 14 $kms^{-1}$ and 11 $kms^{-1}$, respectively. Impacts onto Europa's and Ganymede's surfaces made by retrograde dust can be considered to occur at double the Keplerian orbital velocity. Similarly, if there is variation in the orbital velocity of the prograde dust, some impacts into Europa's and Ganymede's surfaces could happen at significantly lower velocities. \\*
 
\noindent\textbf{Fig. \ref{dust}}

\section{Structure of the dust model and calculation of input parameters.}
The dust cloud model used in these studies is based upon previous dust models developed by \citet{Kruger2003} and \citet{Krivov2003}. We implement experimental and modelling impact data to reduce the number of unknown variables such as the slope of the ejecta size distribution and the size-velocity relation in ejected fragments. The relationship between the influx and outflux is determined through the mass yield \citep{KoschnyandGrun2001}. The outflux populates the space around Europa or Ganymede, where part of the ejecta escapes (unbound dust), and part falls back to the surface (bound dust). In order to compare the contributions from different influx populations to the dust around Europa or Ganymede, outfluxes caused by each influx population were calculated separately and then combined. The model predicts the dust densities of the ejecta outflux at any altitude within Europa's or Ganymede's Hill sphere where these dust densities can be presented as a size distributions of dust. Essentially, the model predicts the radial dust arising from the surfaces and can predict the size distribution of dust fragments in any 'onion-shell' around Europa or Ganymede. \\*
For each micrometeoroid impact, total ejecta mass is calculated (subsection 3.1); then the total ejecta is transformed into ejecta size distribution (subsection 3.2); and each ejecta fragment is allocated its corresponding ejecta speed, based on its size (subsection 3.3). Altogether combined, the total ejecta dust outflux from the surfaces of Europa and Ganymede are calculated (section 4) and verified against existing data (section 5). 

\subsection{The ejecta mass yield and the total ejecta mass.}
The ejecta yield, $Y$, defined as the ratio of the total ejected mass, $m_c$, to the mass of the impactor, $m_p$, shown in Eq. \ref{yield}, represents the efficiency of the material ejection in an impact event \citep{KoschnyandGrun2001}. For the same impact conditions, the excavated crater volume in ice-silicate target decreases with increasing silicate content, $S$ \citep{LangeandAhrens1987, KoschnyandGrun2001, Hiraoka2008}. High $S$ values correspond to the non-ice material and $S$=0 to pure ice. 
\begin{equation}
\label{yield}
Y=m_c/m_p=V_{1,0}(V_{1,100}/V_{1,0})^{S/100}\rho{_t}2^{-b}m_p^{b-1}\upsilon_p^{2b}
\end{equation}
$V_{1,0}$=6.69$\times{10^{-8}}$ $m^3J^{-1}$ and $V_{1,100}$=$10^{-9}$ $m^3J^{-1}$ are crater volumes made by an impact KE of 1 $J$, for $S$=0\% and $S$=100\% silicate content, respectively, and $\rho{_t}$ the target density in $g\,cm^{-3}$, $m_p$ and $\upsilon{_p}$ are the mass and velocity of the impactor in $g$ and $kms^{-1}$, respectively, and $b$=1.23. The density of the target for different values of $S$ is calculated using the linear mixing model \citep{KoschnyandGrun2001} shown in Eq. \ref{mixing} ($\rho_{ice}$=0.927 $gcm^{-3}$ and silicate, $\rho_{sil.}$=2.8 $gcm^{-3}$).
\begin{equation}
\label{mixing}
1/\rho{_t} =(1-S)/\rho_{ice}+S/\rho_{sil.}
\end{equation}
Europa's and Ganymede's surfaces are mostly covered with water ice, but contain some surface contaminants that are unevenly distributed. Europa is a bright icy moon with a young-looking surface and a high albedo. In previous studies of the dust cloud around Europa \citet{Kruger2003} used $S$=0, but here we consider the non-icey part to have a non-negligible effect on the ejecta, hence in Eq. \ref{mixing} for Europa $S$=0.1. This should account for some amounts of non-ice material observed in spectra of Europa's surface, both from Earth (e.g. \citet{Spencer2006}) and in Galileo flybys \citep{McCord1998}. From global geological mapping of Ganymede, it is estimated that about 2/3 of its surface is covered with a dark non-ice material \citep{Patterson2007}, hence we assume that $S$=0.3 for Ganymede. This value is also in agreement with \citet{Krivov2003} who used the same value for their dust cloud around Ganymede calculations, but based the approximation on albedo measurements. Both on Europa and Ganymede the dark non-ice material is mostly found in topographical lows. The mass yield, $Y$, and total ejected mass, $m_c$, for all impactor dust grains that bombard Europa and Ganymede were calculated using Eq. \ref{yield}. \citet{Kruger2003} used a constant value $Y$=$10^4$, whereas in our model we use variable mass yields depending on impactor properties.

\subsection{The cumulative ejecta size distribution of the ejected fragments.}
According to the fragmentation law, the number of ejected fragments during a cratering event can be characterized by a power law function of the fragment's size. Therefore, the cumulative ejecta size distribution is represented in the form $N=Ad^{-B}$, where $N$ is the number of fragments larger than the fragment size, $d$. From impact experiments onto Europa surface analogue materials, \citet{Miljkovic2011a} suggested that impact ejecta from both pure ice and hydrated minerals fragment in a similar way, and that the fragmentation pattern is independent of impact conditions (such as impact angle). Assuming that these distributions are independent of the impactor size, velocity and incidence angle \citep{Miljkovic2009}, they can be applied to other impact events, such as dust clouds around Europa and Ganymede. In these previous experiments, the slope of the cumulative ejecta size distribution was $B$=1.5 \citep{Miljkovic2011a} for both ice and non-ice materials. Assuming that these materials are suitable surface analogues, the results can be applied over the whole surface of Europa and Ganymede. \\*
The distribution coefficient, $A$, depends on the total impact ejecta mass (Eq. \ref{ahrens}), but also on the target material properties: the largest ejected fragment, $m_b$, that represents 1\% of the total ejecta, $M_T$, in each impact and the ejecta fragments' size distribution slope, $B$ \citep{OKeefeandAhrens1985}. 
\begin{equation}
\label{ahrens}
A=M_T \frac{1-B}{Bm_b^{1-B}}
\end{equation}
The cut-off for the smallest ejecta fragment size is set by the current laboratory detection capability ($10^{-15}g$, S. Barber, S. Sheridan, priv. comm.). The Io population was subsequently neglected as the largest ejecta fragment would be smaller than $10^{-15}$g.

\subsection{The velocity distribution of ejected fragments.}
\begin{figure}[htb]
\begin{center}
\includegraphics[scale=0.6]{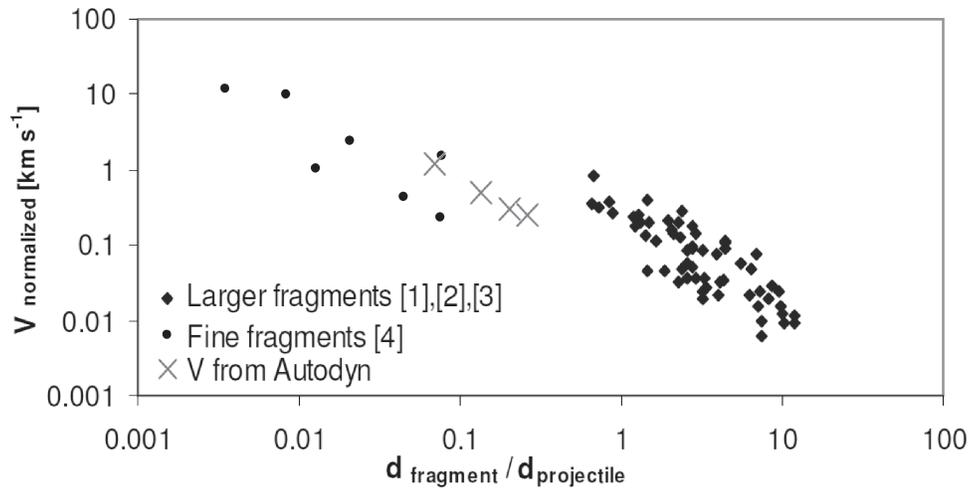}
\end{center}
\caption{Ejecta fragment velocity distribution normalized to $\sqrt{\sigma _t / \rho _t}$, where $\sigma _t$ is the impacted surface strength and $\rho _t$ surface density, calculated in ANSYS AUTODYN-2D impact simulations and compared to similar ejecta fragmentation experiments by: [1] \citet{NakamuraFujiwara1991}, [2] \citet{Nakamura1993}, [3] \citet{FujiwaraTsukamoto1980}, [4] \citet{Nakamura1994}, provides a good fit for $c_1$ and $c_2$ distribution coefficients, for the ejecta size-velocity relation \citep{Melosh1984}.}
\label{SPHvel}
\end{figure}

Each of the ejected fragments was assigned a velocity calculated using Eq. \ref{velocity} that corresponds to the ejecta size-velocity relation in \citet{Melosh1984}, where $\upsilon$ is the initial velocity of an ejected fragment, $\sigma_{t}$ and $\rho_{t}$ are the surface tensile strength and density, respectively. 
\begin{equation}
\label{velocity}
\upsilon/\sqrt{\sigma_{t}/\rho_{t}}=c_1(d/d_{proj})^{c_2}
\end{equation}
The coefficient, $c_1$, and exponent, $c_2$, were derived from impact modelling and impact experiments into water ice and non-ice brittle materials \citep{Miljkovic2010} and are equal to $c_1$=0.06 and $c_2$=-1.16. Details of numerical impact simulations are shown in \citet{Miljkovic2010}. In summary, we used ANSYS AUTODYN-2D SPH (Smooth Particle Hydrodynamic) shock physics code to model impacts into ice and non-ice brittle rocky material. The ejecta fragments were resolved as clusters of SPH cells, from which their size distribution relative to the size of the largest fragment was measured and their velocity recorded. For accuracy, the fragments were distributed in mass bins, each represented by X in Fig. \ref{SPHvel}. This size-velocity distribution was verified against the experimental impact ejecta fragmentation study by \citet{Miljkovic2011b} to validate the numerical simulation results. By fitting Eq. \ref{velocity} to the data in Fig. \ref{SPHvel} we derived the coefficients $c_1$ and $c_2$. Fig. \ref{SPHvel} also shows the ejecta fragment size-velocity distribution from similar impact experiments by \citet{NakamuraFujiwara1991, Nakamura1993, FujiwaraTsukamoto1980, Nakamura1994}.

\textbf{Fig. \ref{SPHvel}}

From the perspective of impacting micrometeoroids, the top mm surface layer at Europa and Ganymede is assumed to resemble snow \citep{HansenandMcCord2004, Schenk2007}, so its strength can be expected to be lower than in the case of non-porous ice. We use $\sigma_t$=0.3 MPa, corresponding to the approximate strength of snow \citep{Petrovic2003}. The density, calculated from the linear mixing model (Eq. \ref{mixing}) was 0.99 $gcm^{-3}$ for Europa (10\% silicate and 90\% ice \citep{Krivov2003}) and 1.82 $gcm^{-3}$ for Ganymede (70\% silica and 30\% ice \citep{Murchie1992} was 1.55); $d$ and $d_{proj}$ are the size of the ejected fragments and the size of the projectile, respectively. The combination of ejecta fragment mass and corresponding ejecta velocity provides a good estimate of the dust outflux around Europa and Ganymede based on the modelled influx.

\section{Results.}
\subsection{The ejecta outflux from Europa's and Ganymede's surface.}
\begin{figure}[htb]
\begin{center}
\includegraphics[scale=0.85]{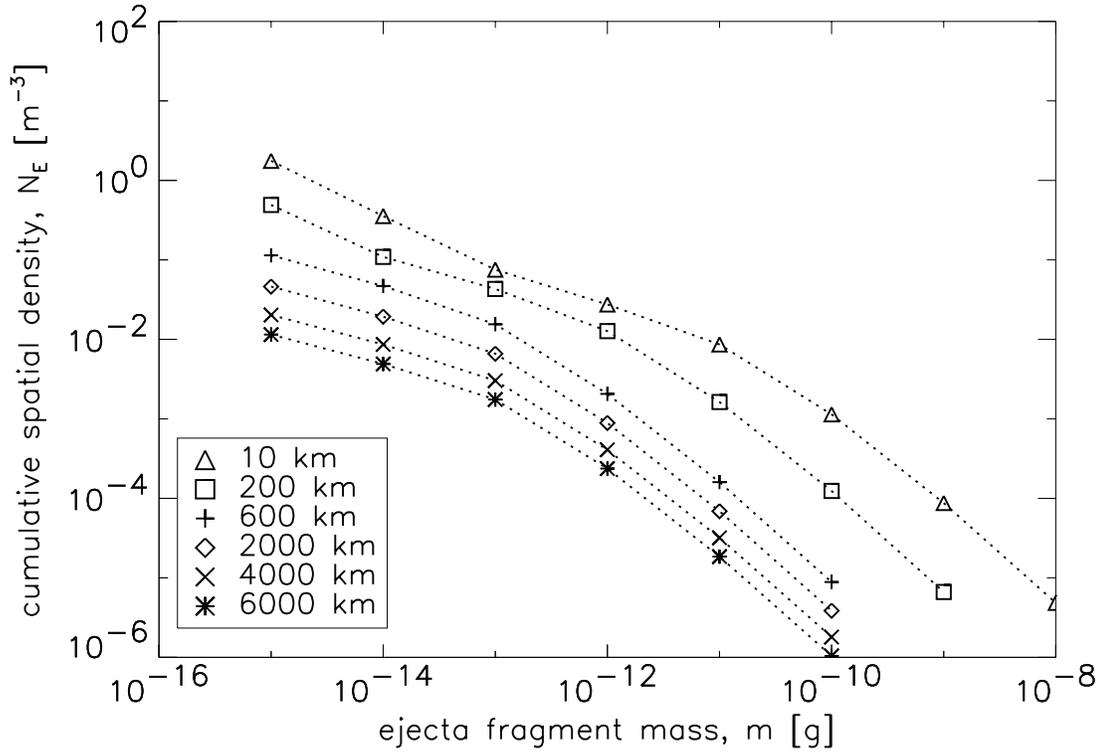}
\end{center}
\caption{Cumulative spatial mass distribution of the ejected surface fragments at different altitudes above Europa's surface shows that the spatial dust density decreases with altitude and there are more smaller than larger fragments, which is both according to the fragmentation law and obeying the gravitational influence on ejecta fragments. Dust masses between $10^{-15}$g and $10^{-11}$g correspond to dust sized from 1 $\mu m$ to 10 $\mu m$.}
\label{Cumm_E} 
\end{figure}

\begin{figure}[htb]
\begin{center}
\includegraphics[scale=0.85]{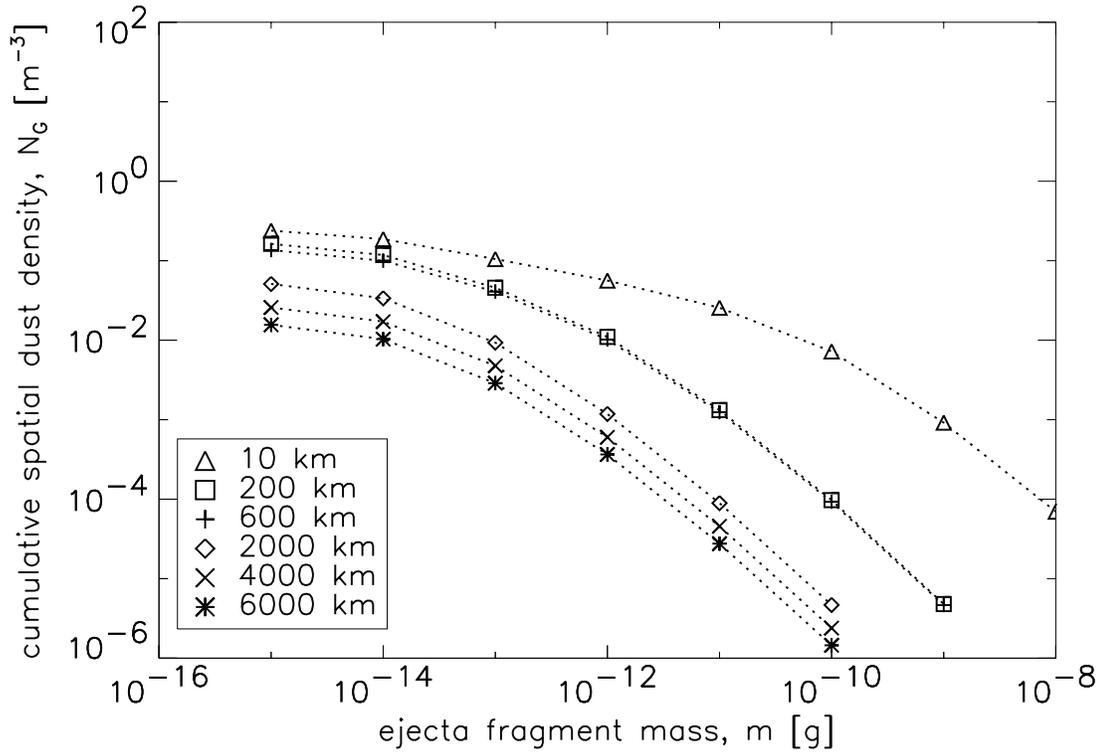}
\end{center}
\caption{Cumulative spatial mass distribution of the ejected surface fragments at different altitudes above Ganymede's surface shows similar trends to Fig \ref{Cumm_E}. Dust masses between $10^{-15}$g and $10^{-11}$g correspond to dust sized from 1 $\mu m$ to 10 $\mu m$.}
\label{Cumm_G} 
\end{figure}

For each impacting dust grain, the total ejecta mass, $m_c$ ($m_c \sim d^{-3}$), is calculated using Eq. \ref{yield}, which is then transformed into an ejecta size distribution in the form of $N=Ad^{-B}$. Coefficient $A$ is calculated using Eq. \ref{ahrens} and $B$ is the experimentally derived distribution slope. For each ejected fragment, the corresponding ejecta size-velocity relation (Eq. \ref{velocity}) is applied and the ejected dust flux leaving the surface can be calculated. The sum of all ejected dust fragments created by the impactor influx, $N$, over the whole surface of a moon, corresponds to the total outflux, $F_{out}(0)$, from the surface ($h$=0) every second. \\*
Within the Hill sphere, the gravitational influence of the moon is the dominant force on ejected dust fragments. Using the energy conservation law (Eq. \ref{energylaw}) between the kinetic and gravitational potential energy of the ejected dust fragments, the speed of the ejecta fragments can be calculated at any altitude. 
\begin{equation}
\label{energylaw}
-\int_{R_m}^{R_m+h} GM d^3 dr/r=\int_0^h d^3\upsilon_{dust}^2(h)/2
\end{equation}
$d$ and $\upsilon_{dust}(h)$ are the size and velocity of an ejected fragment at altitude $h$; $G$ and $R_m$ are the gravitational constant and radius of Europa or Ganymede. By knowing the ejecta velocities at different altitudes, $\upsilon_{dust}(h)$ and the dust outflux ejected from the surface, $F_{out}(0)$, the spatial density at any altitude above the surface, $N(h)$ (in $m^{-3}$), can be calculated using Eq. \ref{onionshell}. The spatial density was calculated using the ejected flux expanding through a sphere of a radius equal to the radial distance from the centre of the moon $(R_m+h)$. 
\begin{equation}
\label{onionshell}
N(h)=F_{out}(0)R_m^2/((R_m+h)^2 \upsilon_{dust}(h))
\end{equation}
Our calculations show of all five populations considered as influxes, only the asteroidal and halo populations are able to create micron size ejecta with sufficient speed to populate orbital altitudes around Europa and Ganymede. The other three (ISD, Io stream dust and ring dust) are found to create a small amount of ejecta that is either too small or too slow, so they have been excluded from models. \\*
The cumulative spatial densities of ejected dust in the form of a mass distribution at different altitudes above the surface were calculated using Eq. \ref{onionshell}. These are shown in Fig. \ref{Cumm_E} for Europa and Fig. \ref{Cumm_G} for Ganymede. The spatial density in both cases drops with radial distance. Ganymede's gravity is nearly twice the gravity of Europa which is reflected in the dust densities. We expect bound ejecta to eventually fall back onto the surface, effectively contributing to the dust population both on its upward and downward trajectory. We do not consider secondary ejecta due to the fact that bound ejecta moves at up to 2-3 $kms^{-1}$ (which corresponds to the escape velocity). 

\noindent\textbf{Figure \ref{Cumm_E}} \\
\noindent\textbf{Figure \ref{Cumm_G}} 

If the kinetic energy ($KE$) of an ejected fragment is greater than the gravitational potential energy ($PE$), then the ejected fragment will escape the gravity of the moon becoming an unbound fragment. Otherwise, the bound fragments would fall back to the surface. The semi-major axis of a fragment's bound orbit can be estimated using $a=-GM_m/(2E)$, where $E=KE-PE$.
 
\begin{figure}[htb]
\begin{center}
\includegraphics[scale=0.85]{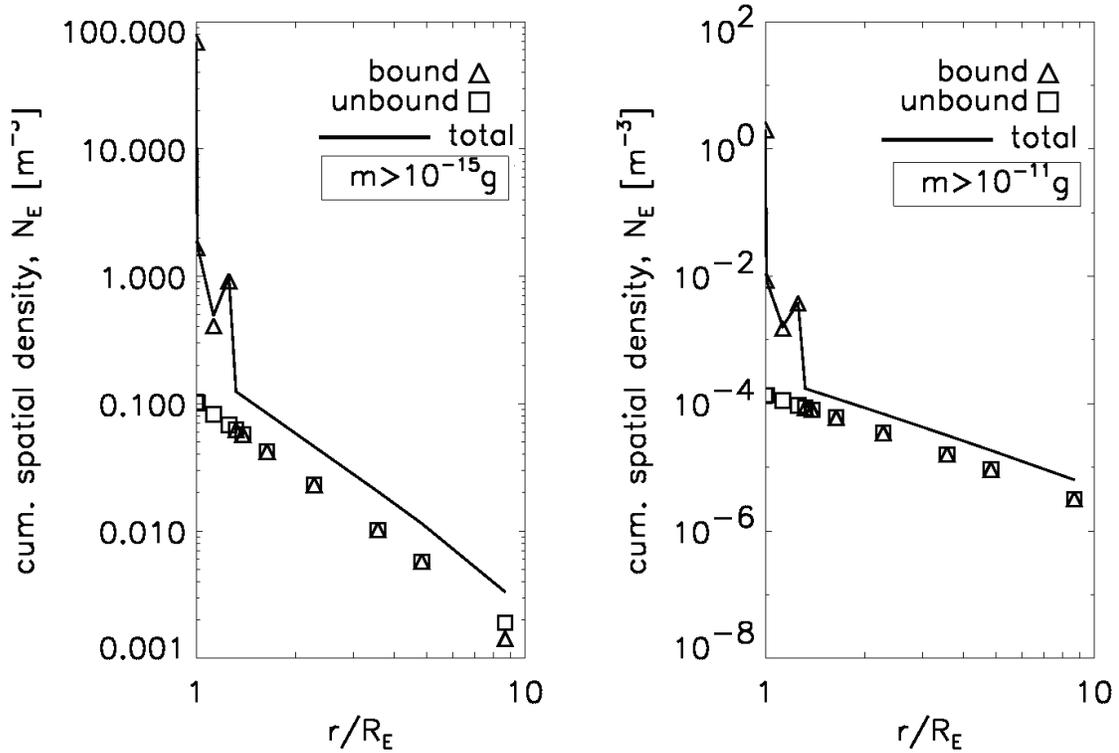}
\end{center}
\caption{a and b. Cumulative spatial density of dust around Europa shown as a function of distance, $r$, from the surface ($R_E$ is radius of Europa), where dust masses are larger than $10^{-15}$ g and $10^{-11}$ g, respectively. Triangles show bound dust, squares show the escaping (unbound) dust and the solid line shows the total ejected dust fragments. Radial distance, $r$, is measured from the center of Europa in $R_E$. Prominent increase in dust density at 0.3 $R_E$ is caused by bound dust reaching its maximum height, slowing down and reversing trajectory in that altitude range.}
\label{Cumr_E} 
\end{figure}

\begin{figure}[htb]
\begin{center}
\includegraphics[scale=0.85]{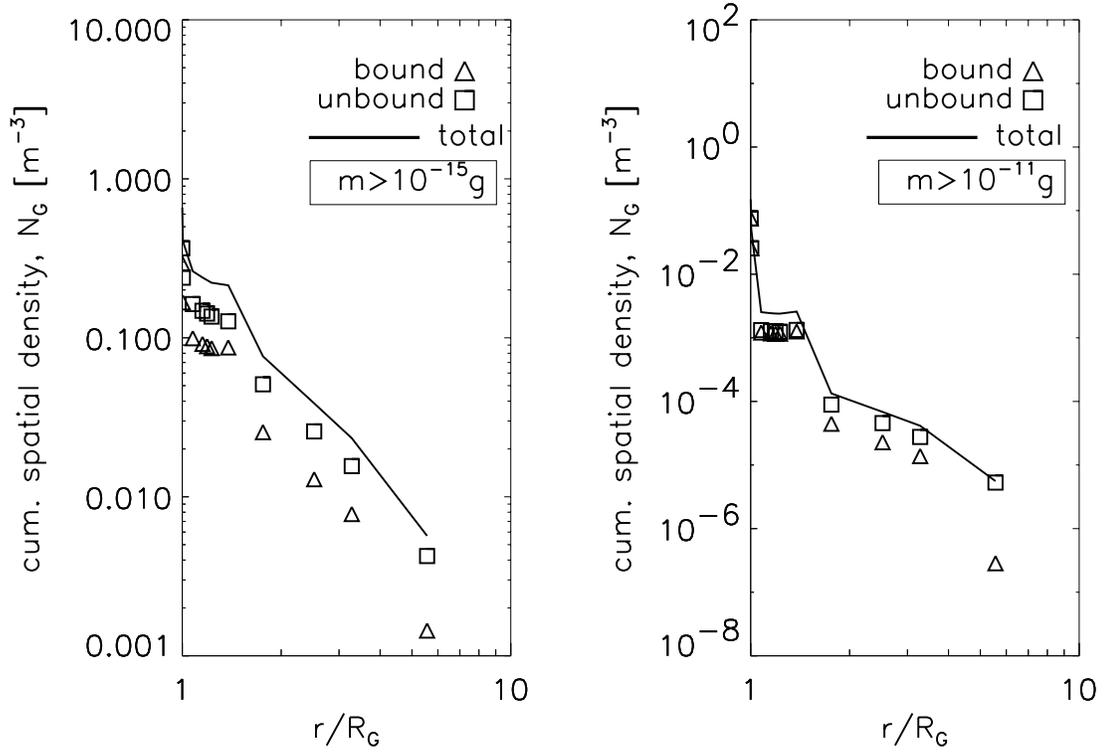}
\end{center}
\caption{a and b. Cumulative spatial density of dust around Ganymede shown as a function of distance, $r$, from the surface ($R_G$ is radius of Ganymede), where dust masses are larger than $10^{-15}$ g and $10^{-11}$ g, respectively. Triangles show bound dust, squares show the escaping (unbound) dust and the solid line shows the total ejected dust fragments. Radial distance, $r$, is measured from the center of Ganymede in $R_G$. An increase in dust density up to 0.4 $R_E$ is caused by bound dust reaching its maximum height, slowing down and reversing trajectory in that altitude range.}
\label{Cumr_G} 
\end{figure}

\noindent\textbf{Figures \ref{Cumr_E}a and b} \\
\noindent\textbf{Figures \ref{Cumr_G}a and b} 

Figs. \ref{Cumr_E}a and \ref{Cumr_G}a show the cumulative spatial density of ejected fragments larger than $10^{-15}$ g (which is considered to be the threshold for mass detection in the laboratory) and Figs. \ref{Cumr_E}b and \ref{Cumr_G}b show fragments larger than $10^{-11}$ g (the mass detection threshold in the Galileo DDS instrument) for Europa and Ganymede, respectively. Due to bound dust's upward and downward trajectory on which dust changes direction and falls back towards the surface, there is a prominent increase in dust density at around 500 km (0.3 $R_E$) at Europa and a smaller increase around Ganymede at about 1000 km (0.4 $R_G$). Assuming that the top surface layer strength is similar on Europa and Ganymede (0.3 MPa, which corresponds to the strength of snow \citep{Petrovic2003}), due to the higher gravity on Ganymede the distribution between bound and unbound fragments is different. On Europa 99\% of ejected fragments belong to the bound population and will fall back to the surface after reaching the maximum height, whereas on Ganymede 80\% of the ejected fragments are bound. A spacecraft in orbit around Europa or Ganymede could have orbital altitudes down to 200 km, hence it is important to study the dust populations at these altitudes to predict the dust count and size distributions that can be collected and chemically analysed by a dust detector on such an orbiter. 

\section{Verification of the model.}
This model was built on previous models upgraded with available impact experimental and modelling data. The model can be verified against the small amount of observational data collected by Galileo DDS (Dust Detector System) as only a handful of dust grains was measured in its short duration flybys. \\*
During 8 Europa flybys, Galileo DDS collected about 50 grains \citep{Kruger2003} and during 4 Ganymede flybys about 40 grains \citep{Kruger2000} that were believed to come from the dust clouds around the moons ($N_{obs.}$). The available data were transformed into the cumulative spatial distribution, $N=N_E$ or $N_G$ in units of $m^{-3}$, by dividing the data by the spacecraft speed, $\upsilon_{sp.}$=2-9 $kms^{-1}$ at Europa and 2-12 $kms^{-1}$ at Ganymede, the spin-averaged sensor area during the flybys ($A=$0.061-0.0235 $m^2$), the average time each flyby lasted namely about $t$=2 h and the number of flybys, $N_{flybys}$ \citep{Kruger2000, Kruger2003}  (Eq. \ref{Eq:Count}).

\begin{equation}
\label{Eq:Count}
N=\frac{N_{obs.}}{\upsilon _{sp.}tN_{flybys}}
\end{equation}

\begin{figure}[htb]
\begin{center}
\includegraphics[scale=0.85]{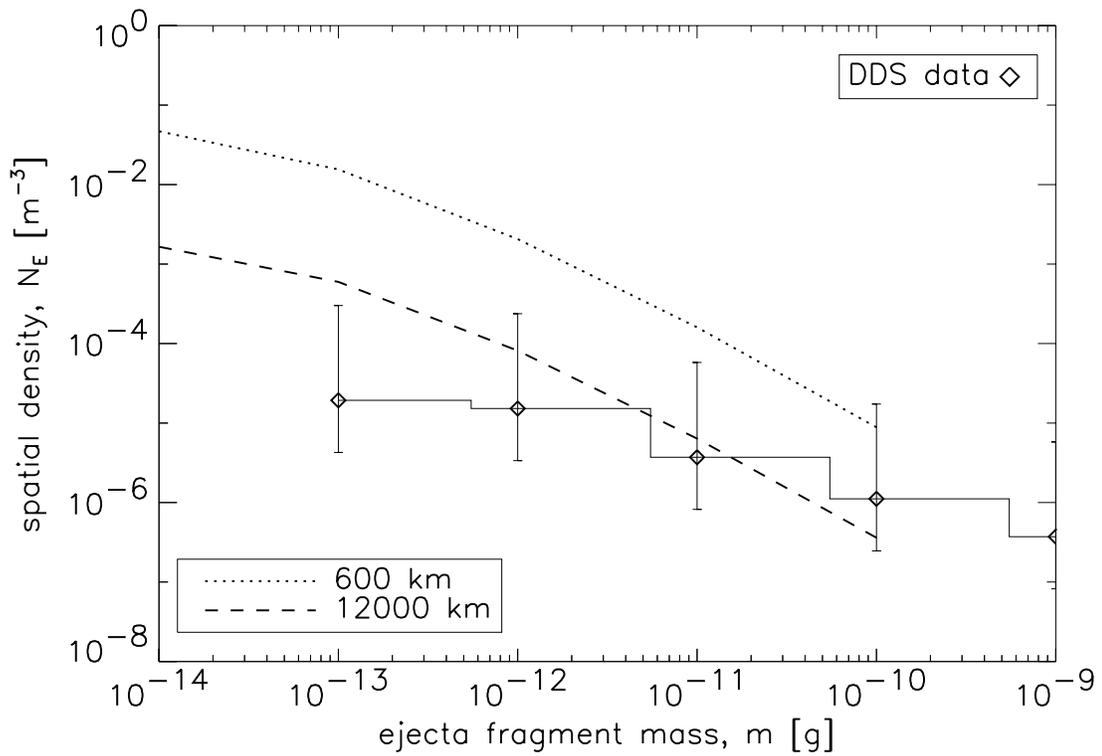}
\end{center}
\caption{Cumulative spatial density of dust around Europa, at 600 km (0.4 $R_E$) and 12000 km (8 $R_E$) altitudes compared with the Galileo DDS data reported in \citep{Kruger2000, Kruger2003}, where flybys were\textsl{•} no closer than 600 km.}
\label{DDS_E} 
\end{figure}

\begin{figure}[htb]
\begin{center}
\includegraphics[scale=0.85]{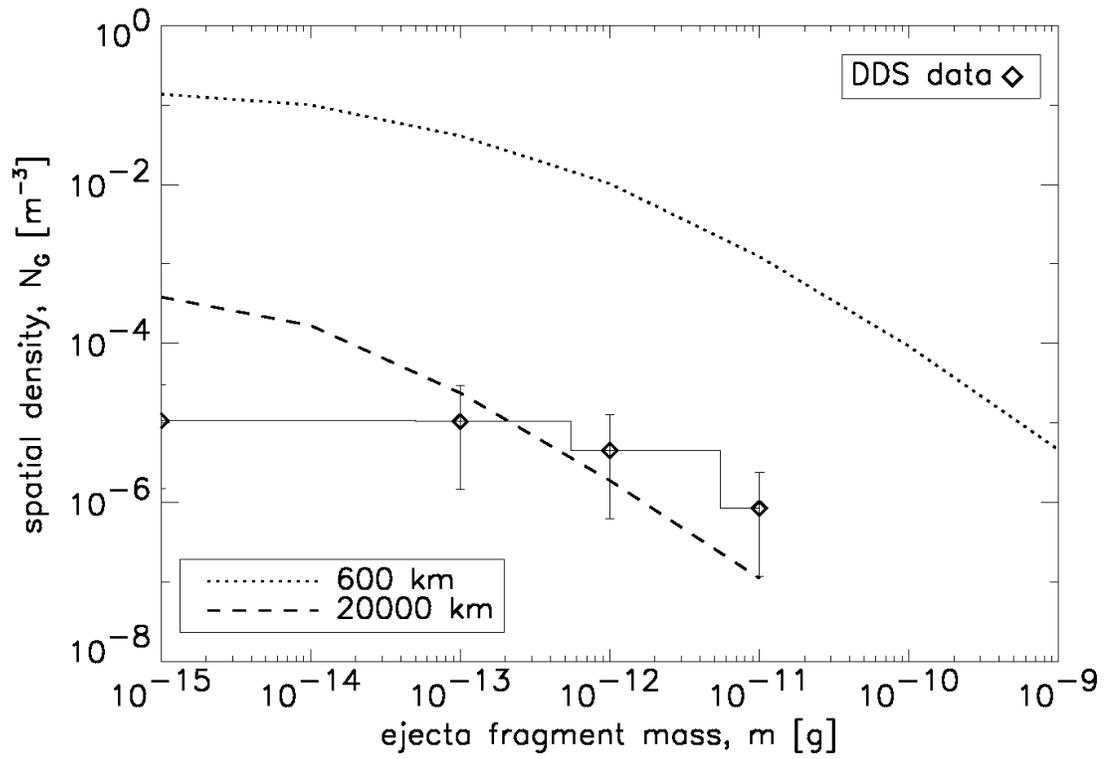}
\end{center}
\caption{Cumulative spatial density of dust around Ganymede, at 600 km (0.2 $R_G$) and 20000 km (8 $R_G$) altitudes compared with the Galileo DDS data reported in \citep{Kruger2000, Kruger2003}, where the flybys ranged between 0.1 and 10 $R_G$.}
\label{DDS_G} 
\end{figure}

Fig. \ref{DDS_E} shows the cumulative spatial density of dust around Europa, at 600 km (0.4 $R_E$) and 12000 km (8 $R_E$) altitude, calculated by our model, compared with the Galileo DDS data reported in \citet{Kruger2000, Kruger2003}, where flybys were no closer than 600 km. Fig. \ref{DDS_G} shows the cumulative spatial density of dust around Ganymede, at 600 km (0.2 $R_G$) and 20000 km (8 $R_G$) altitudes compared with the altitudes over which Galileo collected the data reported in \citet{Kruger2000, Kruger2003}. Galileo flybys ranged between 0.1 and 10 $R_G$ at Ganymede. Figs. \ref{DDS_E} and \ref{DDS_G} show our dust model results are consistent with the Galileo data at the corresponding detection threshold \citep{Grun1992}, apart from the small inconsistency among smaller fragments, which are the most difficult to detect. The vertical error bar in Figs. \ref{DDS_E} and \ref{DDS_G} comes from the variations in the range of flyby velocities, the sensor area and flyby durations. The X-axis shows dust mass in bins.

\noindent\textbf{Figure \ref{DDS_E}} \\*
\noindent\textbf{Figure \ref{DDS_G}}

\begin{figure}[htb]
\begin{center}
\includegraphics[scale=0.5]{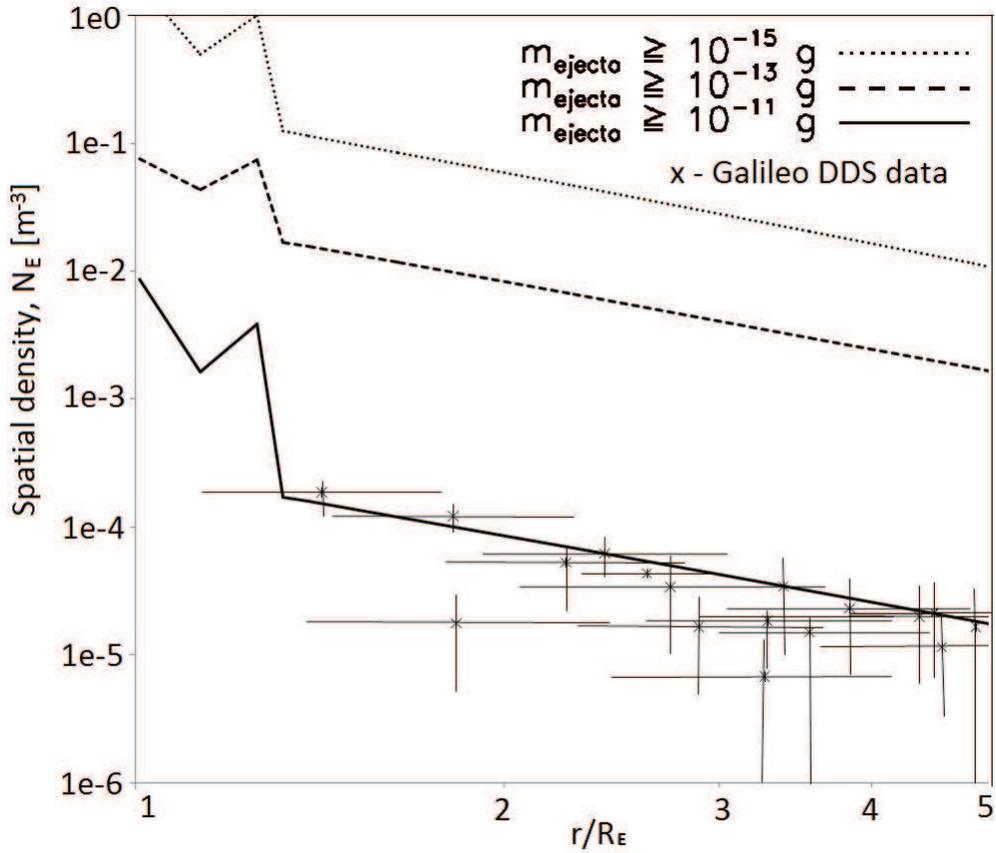}
\end{center}
\caption{Radial spatial density of ejecta from Europa (measured from the centre of the moon), for dust masses larger than three different mass thresholds ($10^{-15}$ g, $10^{-13}$ g and $10^{-11}$ g), compared with Galileo data. Galileo DDS had a detection mass threshold at $10^{-11}$ g for this set of data \citep{Grun1992}, which matches our model predictions.}
\label{Gal_rad_E} 
\end{figure}

\begin{figure}[htb]
\begin{center}
\includegraphics[scale=0.5]{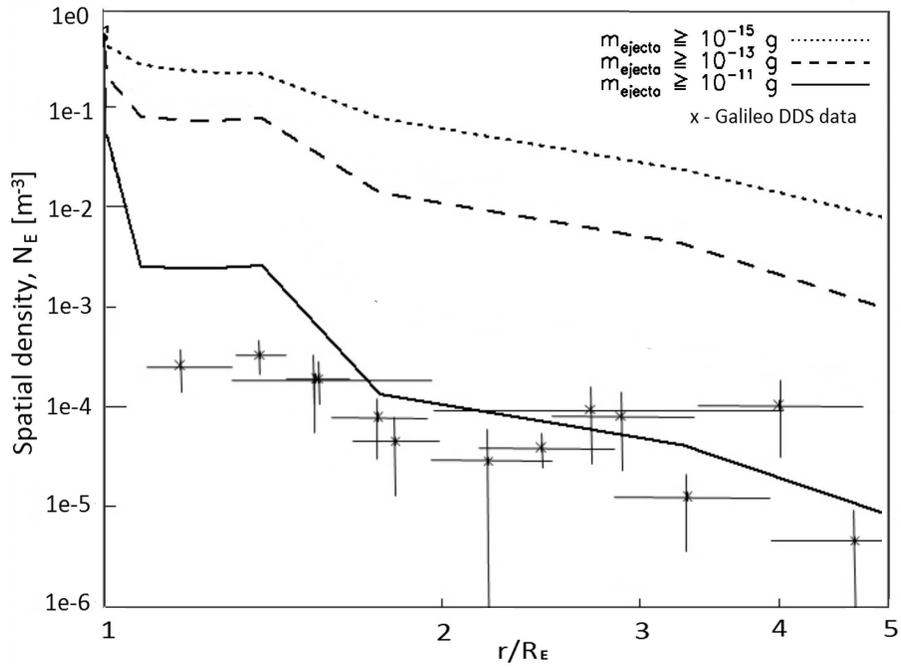}
\end{center}
\caption{Radial spatial density of ejecta from Ganymede (measured from the surface of the moon), for dust masses larger than three different thresholds ($10^{-15}$ g, $10^{-13}$ g and $10^{-11}$ g), compared with Galileo data. Galileo DDS had a detection mass threshold at $10^{-11}$ g for this set of data \citep{Grun1992}, which matches our model predictions, similarly to Fig. \ref{Gal_rad_E}.}
\label{Gal_rad_G}
\end{figure}

Figs. \ref{Gal_rad_E} and \ref{Gal_rad_G} show the spatial dust density at different mass thresholds decreasing with distance, and is also compared with the Galileo DDS data measured at different altitudes from the surfaces of Europa and Ganymede, respectively. The calculated radial spatial density for mass threshold corresponding to Galileo DDS sensitivity threshold ($10^{-11}$ g \citep{Grun1992}) fits the Galileo data collected in Europa and Ganymede flybys. We are confident that the predicted dust densities at lower altitudes, for which there is no observational data, are valid and can provide a confident dust estimate for a dust detector onboard an orbiting spacecraft around Europa or Ganymede.

\noindent\textbf{Figure \ref{Gal_rad_E}} \\*
\noindent\textbf{Figure \ref{Gal_rad_G}} 

\section{Discussion. Significance of input parameters accuracy.}
The most probable input values were used in the dust models presented here. However, the space dust environment is complex and possible input parameters could take a range of values. We investigated the effects a range of input parameter values would have on the outcome of the model. The variations of the ejecta mass yield and surface strength have the largest effect, variation in the slopes of the cumulative ejecta fragment distribution or a change in the largest ejecta fragment affect the dust density to a lesser extent, whereas changing the rest of the input parameters (size-velocity distribution, micrometeoroid impact velocity, fraction of the non-ice content on the surface and micrometeoroid density) affect the model outcome the least. \\*
In our models the largest ejecta fragment was taken to be 1\% of the total ejecta mass, $M_t$, in an impact event. If that value was 0.01\% $M_t$ or 10\% $M_t$, the dust density would increase or decrease, respectively, by a factor of 5. If the ejecta fragmentation slope $B$ changes for $\pm$20\%, the dust density would also change by a factor of 5, moving the distribution more towards larger or smaller fragments, depending upon the slope value.\\*   
The input parameter that has the greatest effect on the model outcome is the strength of the surface material, which is mainly ice. The strength of ice can vary significantly (it depends on temperature, grain size and porosity) from less than 0.3 MPa (corresponding to snow, which is the value used in our model), to 3 MPa for non-porous crystalline ice \citep{Petrovic2003} and up to 10 MPa if the ice contains a fraction of non-ice, silica-like, material \citep{Hiraoka2008}. If the strength is varied between 0.01 and 10 MPa, this would create an order of magnitude (10 times) lower or higher dust density in the dust cloud, respectively. The strength of the surface material on Europa and Ganymede could also vary locally, which is another reason why it is important to emphasize the strength as the most dominant input parameter. The actual density measurements and chemical analysis of dust in the clouds around Europa and Ganymede by a dust detector in orbit would also help narrow down the surface material strength.

\section{Conclusions.}
In this paper, we present an update model for determining the size and spatial density of dust around Europa and Ganymede at any altitude within the moons' Hill sphere, with particular emphasis on the role of micrometeoroid impacts ejecting surface material. The model computes influx distributions from different sources and addresses each ejected fragment individually according to its size and corresponding speed, allowing the complete profile of the dust around Europa and Ganymede to be defined through calculation of the size, velocity and spatial distribution of the complete outflux. \\*
An important feature of this dust model is the ability to predict the dust around Europa for distances closer to the surface than any of the spacecraft flybys made so far as well as for the dust populations below the Galileo DDS detection threshold. We have verified these results by comparing our results with data collected by the Galileo DDS. It is found that the dust spatial densities close to the surface of Europa and Ganymede should be much higher than a simple extrapolation from the actual Galileo data. There should also be much more dust at sizes below the Galileo DDS detection threshold. \\*
This model can be applied to any other atmosphereless body in the Solar System, but above all, it is of great importance for future orbiter missions around Europa and Ganymede, such as recently selected JUICE mission, ESA's next large space mission, to Jupiter System. This study provides scientific support for a dust detector/analyser payload for JUICE or any other future space mission to Europa and/or Ganymede. It also complements surface studies traditionally performed using remote sensing instruments.

\section*{Acknowledgements}
KM thanks Astrium (EADS/UK) for co-funding her doctoral studies, and we thank the anonymous referees for useful comments that helped improve this manuscript. 

\section*{References}
\bibliographystyle{elsarticle-harv}

\newpage
\section*{Figures}

\end{document}